\documentclass[a4paper,11pt]{article}

\usepackage{jheppub}
\usepackage[table]{xcolor}
\usepackage{slashbox}
\allowdisplaybreaks
\usepackage{graphicx}
\usepackage{epstopdf}

\newcommand{\be}{\begin{equation}}
\newcommand{\ee}{\end{equation}}
\newcommand{\bea}{\begin{eqnarray}}
\newcommand{\eea}{\end{eqnarray}}
\newcommand{\nn}{\nonumber}
\newcommand{\bdm}{\begin{displaymath}}
\newcommand{\edm}{\end{displaymath}}

\title{N$^3$LO gravitational quadratic-in-spin interactions at $G^4$} 

\author[a,b]{Mich\`ele Levi,}
\author[a]{Andrew J.~McLeod,}
\author[a]{and Matthew von Hippel}

\affiliation[a]{Niels Bohr International Academy, Niels Bohr Institute,
University of Copenhagen,
\\Blegdamsvej 17, 2100 Copenhagen, Denmark}

\affiliation[b]{Institut de Physique Th\'eorique, 
CEA \& CNRS, Universit\'e Paris-Saclay,
\\ 91191 Gif-sur-Yvette, France}

\emailAdd{michelelevi@nbi.ku.dk}
\emailAdd{amcleod@nbi.ku.dk}
\emailAdd{mvonhippel@nbi.ku.dk}

\abstract{We compute the N$^3$LO gravitational quadratic-in-spin interactions 
at $G^4$ in the post-Newtonian (PN) expansion via the effective field theory (EFT) 
of gravitating spinning objects for the first time. 
This result contributes at the $5$PN order for maximally-spinning compact 
objects, adding the spinning case to the static sector at this PN accuracy. 
This sector requires extending the EFT of a spinning particle beyond linear order in the 
curvature to include higher-order operators quadratic in the curvature that are relevant at 
this PN order. 
We make use of a diagrammatic expansion in the worldline picture, and rely on our recent 
upgrade of the \texttt{EFTofPNG} code, which we further extend to handle this sector. 
Similar to the spin-orbit sector, we find that the contributing three-loop graphs give rise to 
divergences, logarithms, and transcendental numbers. 
However, in this sector all of these features conspire to cancel out from the final result, 
which contains only finite rational terms.} 

\begin{document}

\today

\maketitle

\flushbottom

\section{Introduction} 
\label{intro}

As of 2016 the observation of gravitational waves (GWs) has become a reality 
\cite{Abbott:2016blz}. An unexpected variety of GW events has already been observed from 
black hole and neutron star binary mergers on the first and second observation runs of 
the LIGO and Virgo collaborations \cite{Ligo,Virgo,LIGOScientific:2018mvr}. The continuously 
increasing influx of data expected from the expanding worldwide network of ground-based 
interferometers \cite{Kagra,IndiGO,Punturo:2010zz,PAIK:2016yxn} and anticipated space-based 
detectors in complementary frequencies \cite{Lisa,Luo:2015ght} places significant demands on 
the precision of theoretical predictions. The theoretical waveform templates used for these 
observations are uniquely created using the effective one-body (EOB) framework 
\cite{Buonanno:1998gg,Damour:2012mv}, which in turn relies heavily on the post-Newtonian (PN) 
theory of General Relativity as input \cite{Blanchet:2013haa}. Even relatively high-order 
corrections beyond Newtonian gravity, such as the sixth PN (6PN) order, are required for 
accurate waveforms, and to gain further information about the inner structure of the 
individual components of the binary. 

PN theory provides an analytic treatment of the long inspiral phase, in which the two compact 
objects in the binary move with non-relativistic velocities. The orbital dynamics of the 
compact binaries is an essential ingredient for theoretical waveform models. The current state 
of the art for orbital dynamics for generic compact binaries in PN theory is presented in 
table \ref{stateoftheart}. This table shows the sectors obtained at order 
$n+l+\text{Parity}(l)/2$ in the PN expansion, where $l$ is the order of spin that appears in 
the sector with the corresponding parity for even or odd $l$, and for $l\geq2$ 
finite-size effects have to be tackled. Along the years much of 
the progress in PN theory has been made only for the simple unrealistic case, in which the 
compact objects are not spinning and are considered as point-masses (corresponding to the 
first row in table \ref{stateoftheart}), due to the considerable conceptual and computational 
difficulty of treating spinning objects in gravity. 
\begin{table}[t]
\begin{center}
\begin{tabular}{|l|r|r|r|r|r|r}
\hline
\backslashbox{\quad\boldmath{$l$}}{\boldmath{$n$}} &  (N\boldmath{$^{0}$})LO
& N\boldmath{$^{(1)}$}LO & \boldmath{N$^2$LO}
& \boldmath{N$^3$LO} & \boldmath{N$^4$LO} & \boldmath{N$^5$LO} 
\\
\hline
\boldmath{S$^0$} & 1 & 0  & 3 & 0 & 25 & 0
\\
\hline
\boldmath{S$^1$} & 2 & 7 & 32 & 174 & & 
\\
\hline
\boldmath{S$^2$} & 2 & 2 & \textbf{18} & \textbf{52} & &
\\
\hline
\boldmath{S$^3$} & 4 & \cellcolor[gray]{0.9} \textbf{24} & \cellcolor[gray]{0.9} 
& \cellcolor[gray]{0.9} & \cellcolor[gray]{0.9} & \cellcolor[gray]{0.9}
\\
\hline
\boldmath{S$^4$} & 3 & \textbf{5} \cellcolor[gray]{0.9} & \cellcolor[gray]{0.9} 
& \cellcolor[gray]{0.9} & \cellcolor[gray]{0.9} & \cellcolor[gray]{0.9} \\
\end{tabular}
\caption{
The minimum number of $n$-loop graphs within the EFT framework for 
each PN sector that has been completed to date for the orbital dynamics of 
generic compact binaries. The gray area indicates the sectors that correspond 
to gravitational Compton scattering with quantum spins $s \ge 3/2$.
}
\label{stateoftheart}
\end{center}
\end{table}

Nevertheless, in recent years remarkable progress has also been made in the spinning sectors 
via the effective field theory (EFT) approach to PN gravity 
\cite{Goldberger:2004jt,Levi:2018nxp} within the framework of the EFT of gravitating spinning 
objects \cite{Levi:2015msa}. The filled entries in table \ref{stateoftheart} show the number 
of $n$-loop graphs that contribute within the EFT framework in each N$^n$LO sector that has 
been completed to date. The boldface entries in the table have been completed 
only via the EFT of gravitating spinning objects \cite{Levi:2015msa}, in a series of works 
\cite{Levi:2011eq,Levi:2014sba,Levi:2014gsa,Levi:2015uxa,Levi:2015ixa,Levi:2016ofk,
Levi:2017kzq,Levi:2019kgk,Levi:2020kvb,Levi:2020lfn}. In particular, in \cite{Levi:2015msa} 
the leading non-minimal gravitational couplings to all orders in spin were formulated, thus 
enabling the completion of all sectors with finite-size effects up to the current state of the 
art at the $4$PN order. 

This line of work was recently extended along both of the axes of table \ref{stateoftheart}: 
The first three-loop calculation in the spinning sector was computed in \cite{Levi:2020kvb}, 
using the publicly-available \texttt{EFTofPNG} code \cite{Levi:2017kzq}, and the cubic- and 
quartic-in-spin sectors were tackled at one-loop level in \cite{Levi:2019kgk,Levi:2020lfn}. 
The latter sectors thus uniquely explore the gray area in table \ref{stateoftheart}, which 
corresponds to gravitational Compton scattering with quantum spins $s \ge 3/2$, as sectors 
with classical spin at the $l$-th order correspond to scattering with quantum spin of $s=l/2$ 
\cite{Arkani-Hamed:2017jhn}. Altogether, this has been pushing the precision frontier to the 
$5$PN order.

This work derives for the first time the complete N$^3$LO gravitational interactions which are 
quadratic in the spins from a diagrammatic expansion at order $G^4$. All contributions at this 
order are static, meaning that they involve no explicit factors of the velocities of the 
compact objects. This sector involves both three-loop graphs and finite-size effects due to 
the spin-induced quadrupole. It enters at the $5$PN order for maximally-spinning compact 
objects, thus further pushing the precision frontier of PN gravity. This work builds on the 
EFT of gravitating spinning objects \cite{Levi:2015msa} and its implementation in the 
publicly-available \texttt{EFTofPNG} code in \cite{Levi:2017kzq}, as well as on their recent 
implementations and upgrade at the two-loop level \cite{Levi:2015ixa} and at the three-loop 
level \cite{Levi:2020kvb}. 

As noted above the spinning sectors are considerably more challenging to handle than the 
non-spinning ones. First, contrary to the non-spinning case, in the spinning sectors all 
possible graph topologies are realized at each order of $G$ \cite{Levi:2008nh}. While there 
are no three-loop graphs that enter in the non-spinning N$^3$LO sector, in the present sector 
there are $52$ such graphs (see table \ref{stateoftheart}), which turn out to be precisely 
those that produce divergences and logarithms. Further, this sector contains $163$ graphs to 
evaluate compared to only $8$ in the N$^3$LO sector without spins. We also recall that spins 
are derivatively coupled, which in this sector with two spins leads to integrand tensor 
numerators up to rank eight, comparable to N$^5$LO in the sector without spins. Finally, at 
this order we also have to consider further contributions from higher-order operators with 
spins that are beyond linear in the curvature (note, however, that tails that involve spin 
couplings do not contribute at the order considered here 
\cite{Blanchet:2011zv,Blanchet:2013haa}). 

The paper is organized as follows. We begin in section \ref{theoryquadspin} by reviewing the 
formal framework of the EFT of gravitating spinning objects, which we extend in section 
\ref{moretheory} to include further non-minimal couplings in the effective action of a 
spinning particle that become relevant at this order. We proceed in section \ref{bringiton} to 
consider the diagrammatic expansion of interactions that are quadratic in the spins, 
highlighting how the graphs in this sector can be constructed out of the graphs in lower-order 
sectors. We discuss the total outcome for the sector in section \ref{result}, and conclude in 
section \ref{theendmyfriend}.

\section{EFT of gravitating spinning objects}
\label{theoryquadspin}

We start by reviewing the framework of the EFT of gravitating spinning objects, which enables 
our derivation of the N$^3$LO quadratic-in-spin sectors from a diagrammatic expansion at order 
$G^4$. We follow closely the setup presented in the recent work \cite{Levi:2020kvb} and also 
build on \cite{Levi:2015ixa}, keeping similar conventions and notations. First we will extend 
the one-particle effective action that was required for the N$^3$LO spin-orbit in 
\cite{Levi:2020kvb}, introducing further Feynman rules that are required in this sector. Thus, 
from the two-particle effective action that describes a compact binary 
\cite{Goldberger:2004jt,Levi:2018nxp}:
\be \label{2ptact}
S_{\text{eff}}=S_{\text{g}}[g_{\mu\nu}]+\sum_{a=1}^{2}S_{\text{pp}}(\lambda_a),
\ee
we will only be concerned with $S_{\text{pp}}$, the worldline action of a spinning particle 
for each of the two components of the binary, where $\lambda_a$ are their respective worldline 
parameters. 

From the pure gravitational action, $S_{\text{g}}$, given in \cite{Levi:2020kvb} in the 
harmonic gauge, no additional ingredients are required beyond those presented in 
\cite{Levi:2020kvb}. Hence, we use similar propagators for the gravitational field components 
in terms of the Kaluza-Klein (KK) decomposition \cite{Kol:2007bc,Kol:2010ze}, and no further 
bulk vertices are required beyond those which were added in \cite{Levi:2020kvb}. As in 
\cite{Levi:2020kvb} the spatial dimension, $d$, is kept generic throughout, which also matches 
what is done in the  \texttt{EFTofPNG} code \cite{Levi:2017kzq}. We use the corresponding 
$d$-dimensional gravitational constant \cite{Levi:2020kvb},
\be
G_d\equiv G_N \left(\sqrt{4\pi e^\gamma} \,R_0 \right)^{d-3},
\ee
where $G_N\equiv G$ is Newton's gravitational constant in three-dimensional space, 
$\gamma$ 
is Euler's constant, and 
$R_0$ is a fixed renormalization scale. This corresponds to the modified minimal subtraction 
($\overline{\text{MS}}$) prescription in dimensional regularization used in this work, see 
e.g.~\cite{Peskin:1995ev}. As in the N$^3$LO sector at order $G^4$ in \cite{Levi:2020kvb} 
relativistic corrections to the propagators are not relevant in this paper.

We now focus on the point-particle action for each of the spinning particles 
\cite{Levi:2015msa,Levi:2018nxp}. The quadratic-in-spin sector includes three types of 
interaction: i.~the coupling between the spins of both objects, referred to as the 
spin$_1$-spin$_2$ interaction, ii.~the nonlinear interaction of each object's spin with 
itself, referred to as spin$_1$-spin$_1$ interaction, and iii.~finite-size effects that 
involve the spin-induced quadrupole of each of the rotating objects. Due to the latter 
interaction, the effective action should be considered beyond minimal coupling, which is 
sufficient for the spin-orbit sector, and for interactions that are dependent on the linear 
spin couplings as in the spin$_1$-spin$_2$ and spin$_1$-spin$_1$ interactions. Considering all 
these interactions, the effective action of each of the spinning particles reads 
\cite{Levi:2015msa,Levi:2018nxp}:
\begin{align} \label{spinptact}
S_{\text{pp}}(\lambda)=&\int 
d\lambda\left[-m \sqrt{u^2}-\frac{1}{2} \hat{S}_{\mu\nu} \hat{\Omega}^{\mu\nu}
	-\frac{\hat{S}^{\mu\nu} p_{\nu}}{p^2} \frac{D p_{\mu}}{D \lambda}
+L_{\text{NMC}}
\left[g_{\mu\nu},u^{\mu},S^{\mu}\right]\right],
\end{align}
where $L_{NMC}$ denotes the non-minimal coupling part of the action induced by the spin of the 
object. This part is initially formulated in terms of the definite-parity pseudovector 
$S_{\mu}$, as defined in \cite{Levi:2014gsa,Levi:2015msa,Levi:2019kgk}. Beyond these 
non-minimal couplings all of the previous terms in the action are similar to what is noted in 
\cite{Levi:2020kvb}, where the difference with the action in 
\cite{Hanson:1974qy,Bailey:1975fe,Porto:2005ac} was highlighted. 

The non-minimal coupling part of the effective action involving the spin was constructed in 
\cite{Levi:2015msa}, where the spin-induced non-minimal couplings to all orders in spin and 
linear in the curvature were derived to be:
\bea \label{spinleadingnmc}
L_{\text{NMC(R)}}&
=&\sum_{n=1}^\infty \frac{(-1)^n}{(2n)!}\frac{C_{ES^{2n}}}{m^{2n-1}}
D_{\mu_{2n}}\cdots D_{\mu_{3}}\frac{E_{\mu_{1}\mu_{2}}}{\sqrt{u^2}}
S^{\mu_1}S^{\mu_2}\cdots S^{\mu_{2n-1}}S^{\mu_{2n}}\nn\\
&+&\sum_{n=1}^\infty \frac{(-1)^n}{(2n+1)!}\frac{C_{BS^{2n+1}}}{m^{2n}}
D_{\mu_{2n+1}}\cdots D_{\mu_{3}}\frac{B_{\mu_{1}\mu_{2}}}{\sqrt{u^2}}
S^{\mu_1}S^{\mu_2}\cdots S^{\mu_{2n}}S^{\mu_{2n+1}},
\eea
which involves new spin-induced Wilson coefficients, the electric and magnetic curvature 
components of definite parity, defined as
\bea
E_{\mu\nu}&\equiv& R_{\mu\alpha\nu\beta}u^{\alpha}u^{\beta}, \label{elec}\\
B_{\mu\nu}&\equiv& \frac{1}{2} \epsilon_{\alpha\beta\gamma\mu} 
R^{\alpha\beta}_{\,\,\,\,\,\,\,\delta\nu}u^{\gamma}u^{\delta}\label{mag},
\eea
as well as their covariant derivatives, $D_{\mu}$. Of the infinite series in 
eq.~\eqref{spinleadingnmc} only the first operator is required in this paper, corresponding to 
the spin-induced quadrupole that reads:
\be
L_{ES^2}=-\frac{C_{ES^2}}{2m}\frac{E_{\mu\nu}}{\sqrt{u^2}} S^\mu S^\nu. 
\label{es2}
\ee
Up to missing generic symmetry and construction considerations, a closely related expression 
was used in \cite{Porto:2008jj}, which conformed with the well-known LO spin-squared 
interaction from \cite{Barker:1975ae,Poisson:1997ha}. 

There are no additional Feynman rules (such as worldline mass and spin couplings) required for 
the spin$_1$-spin$_2$ and spin$_1$-spin$_1$ interactions, which originate from the 
minimal-coupling part of the point-particle action, beyond those presented in 
\cite{Levi:2015uxa,Levi:2020kvb}. Thus, we turn now to the Feynman rules required for this 
sector due to the interaction involving the spin-induced quadrupole, which go beyond the rules 
in previously computed lower-order spinning sectors \cite{Levi:2015ixa}. All of the 
aforementioned Feynman rules can be obtained within the public \texttt{EFTofPNG} code 
\cite{Levi:2017kzq} for a generic number of spatial dimensions $d$. The new relevant Feynman 
rules for this sector were obtained by further extending the \texttt{FeynRul} module of the 
\texttt{EFTofPNG} code \cite{Levi:2017kzq}.

For the three-graviton coupling to the worldline spin-induced quadrupole, the new required 
Feynman rule is
\begin{align}
\label{eq:sqphi2sigma} \parbox{12mm}{\includegraphics[scale=0.6]{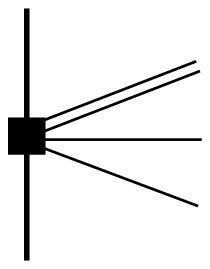}}
& = -\frac{C_{ES^2}}{4m} \frac{1}{d-2} \int dt 
\Bigg[dS^{i}S^{j}
\Big(2\sigma_{ik}\big(
\partial_j \phi \partial_k \phi + \phi \,\partial_j\partial_k\phi \big)       
+ \phi \, \partial_k\phi  \big(2\partial_i\sigma_{jk}-\partial_k\sigma_{ij}\big)\Big)\nn\\
 &\qquad\qquad\qquad\qquad\qquad 
 -S^2\Big(2\sigma_{ij}\big(
 \partial_i\phi \partial_j\phi+ d\phi\,\partial_i\partial_j\phi \big) 
 +d\phi\,\partial_i\phi\big(2\partial_j\sigma_{ij}-\partial_i\sigma_{jj}\big)\Big)\Bigg],
\end{align}
while there is also a new Feynman rule for the four-graviton coupling to the worldline 
spin-induced quadrupole, given by
\begin{align}
\label{eq:sqphi4}  \parbox{12mm}{\includegraphics[scale=0.6]{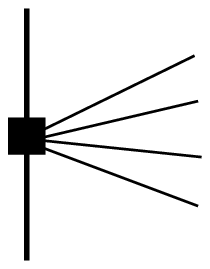}}
 & = \,\frac{C_{ES^2}}{12m} \frac{d^2}{(d-2)^3}\int dt\,\Bigg[
 dS^{i} S^{j} \left(3\phi^2\,\partial_i\phi\,\partial_j\phi
 +\phi^3\,\partial_i\,\partial_j\phi\right)\nn\\
 & \qquad \qquad \qquad \qquad \qquad \quad
 -S^2\left(3\phi^2\left(\partial_i\phi\right)^2
 +d\phi^3\partial_i\partial_i\phi\right)\Bigg].
\end{align}
We remind the reader that these rules are already given in terms of the physical spatial 
components of the local Euclidean spin vector in the canonical gauge \cite{Levi:2015msa}, and 
that all of the indices are in fact Euclidean.

\subsection{Extending the EFT of a spinning particle}
\label{moretheory}

To complete the effective action at this order in PN accuracy in the spinning sectors 
\cite{Levi:2015ixa,Levi:2020kvb} including the current sector, we extend the non-minimal 
coupling part of the effective action of a spinning particle given in 
eq.~\eqref{spinleadingnmc}. This extension contains operators that are quadratic in the 
curvature components and hence stand for tidal deformations of the extended compact 
object. Following similar symmetry considerations and the logic outlined in 
\cite{Levi:2015msa}, we find that the new terms to quadratic order in spin are given by
\begin{align} \label{nmcspinRR}
L_{\text{NMC(R$^2$)}}&
= C_{E^2} \frac{E_{\alpha\beta}E^{\alpha\beta}}{\sqrt{u^2}^{\,3}}
+C_{B^2} \frac{B_{\alpha\beta}B^{\alpha\beta}}{\sqrt{u^2}^{\,3}}+\ldots
\nn\\&
+C_{E^2S^2} S^{\mu} S^{\nu} \frac{E_{\mu\alpha}E_{\nu}^{\,\alpha}}{\sqrt{u^2}^{\,3}}
+C_{B^2S^2} S^{\mu} S^{\nu} \frac{B_{\mu\alpha}B_{\nu}^{\alpha}}{\sqrt{u^2}^{\,3}}
\nn\\&
+C_{\nabla EBS} S^{\mu} \frac{D_{\mu}E_{\alpha\beta}B^{\alpha\beta}}{\sqrt{u^2}^{\,3}}
+C_{E\nabla BS} S^{\mu} \frac{E_{\alpha\beta}D_{\mu}B^{\alpha\beta}}{\sqrt{u^2}^{\,3}}
\nn\\& 
+C_{(\nabla E)^2S^2} S^{\mu} S^{\nu} 
\frac{D_{\mu}E_{\alpha\beta}D_{\nu}E^{\alpha\beta}}{\sqrt{u^2}^{\,3}}
+C_{(\nabla B)^2S^2} S^{\mu} S^{\nu} 
\frac{D_{\mu}B_{\alpha\beta}D_{\nu}B^{\alpha\beta}}{\sqrt{u^2}^{\,3}}+\ldots
\end{align}
These terms involve new tidal Wilson coefficients (that in contrast with 
eq.~\eqref{spinleadingnmc} are defined here to absorb all numerical and mass factors). 

In the first line of eq.~\eqref{nmcspinRR} we have written the leading mass-induced 
quadrupolar tidal deformations, which are known to enter at the 5PN order \cite{Levi:2018nxp}, 
and have suppressed higher-order mass-induced tidal operators, which can be found in 
\cite{Bini:2012gu}. We have additionally shown the adiabatic tidal operators which involve the 
spin up to quadratic order. From dimensional analysis and power-counting considerations (as 
explained in \cite{Levi:2015msa,Levi:2018nxp}), one can infer that the terms in the second 
line of eq.~\eqref{nmcspinRR} enter at the 5PN order, while the terms on the third and fourth 
lines of eq.~\eqref{nmcspinRR} enter only at the 6.5PN and 7PN orders, respectively. 

Therefore at the 5PN order considered in this work there are two new operators which are 
quadratic in the spin and quadratic in the curvature. However, the leading contribution from 
these operators at the 5PN order shows up in the two-graviton exchange topology at order $G^2$ 
(see e.g.~figure 2(a) in \cite{Levi:2020kvb}), and is therefore not relevant in the current 
paper, but rather will be addressed in a subsequent publication. Notice that interestingly if 
this expansion that is quadratic in the curvature is continued to higher orders in spin as in 
\cite{Levi:2020lfn}, only spin orders $l=0,2,4$ that relate to bosons of quantum spin 
$s=0,1,2$ enter at leading order. Spin orders that correspond to fermions and quantum spin 
$s>2$ all require additional covariant derivatives and enter at higher orders.

\section{Perturbative expansion of quadratic-in-spin interactions}
\label{bringiton}

\begin{figure}[t]
\includegraphics[width=0.74\textwidth]{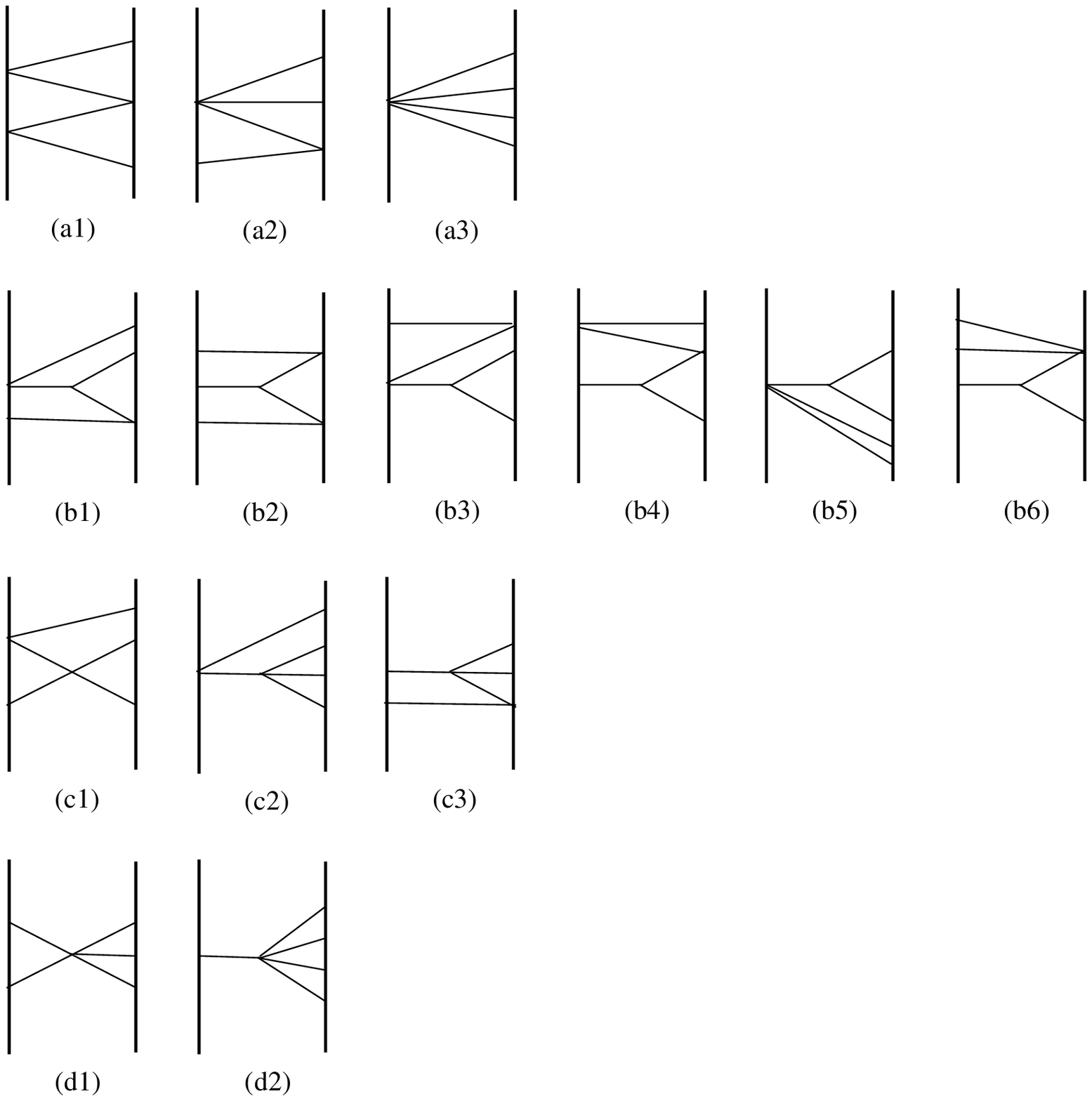}
\includegraphics[width=\textwidth]{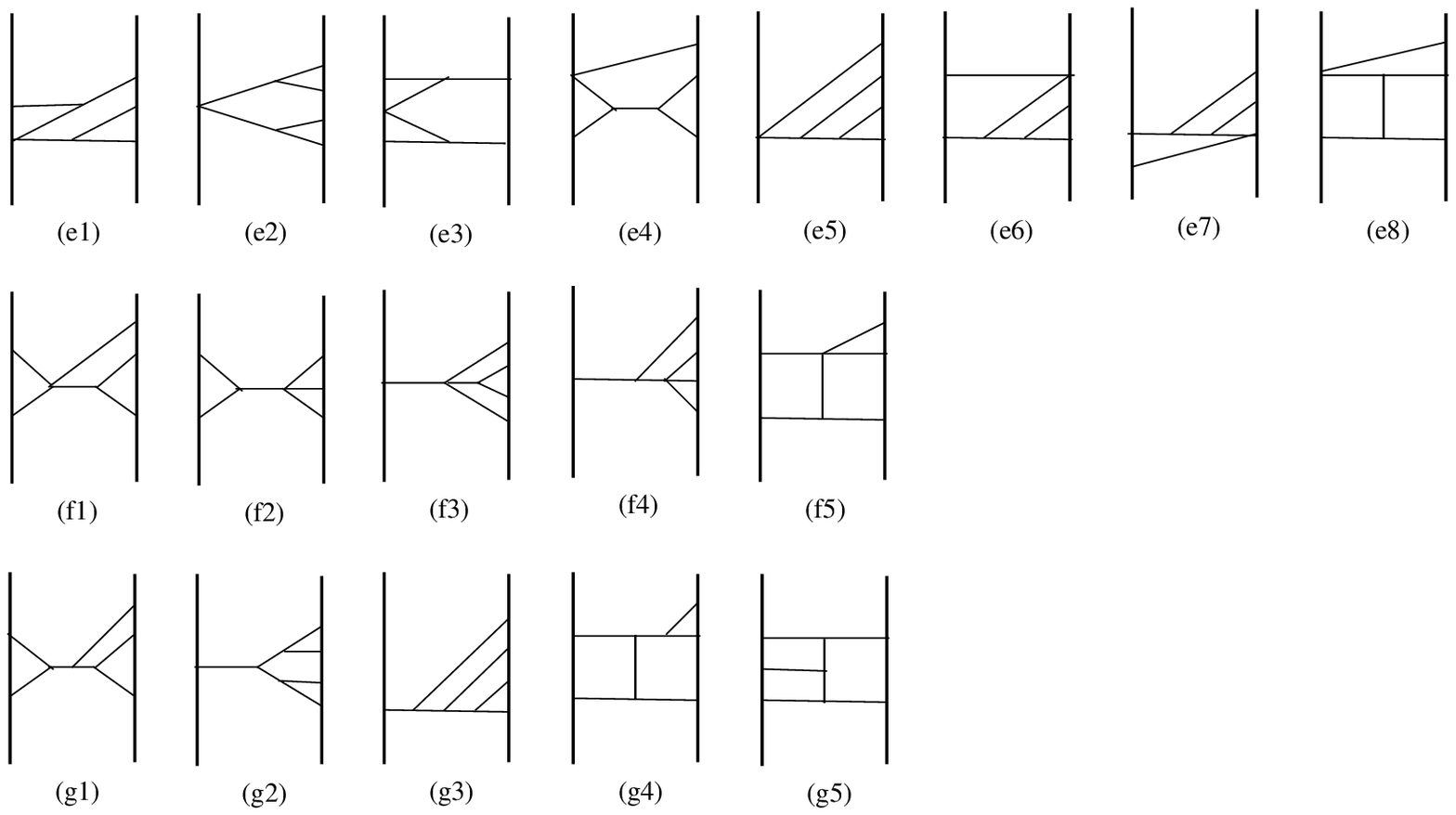}
\caption{Graph topologies at order $G^4$ classified according to their internal bulk vertices 
and the corresponding loop order in the worldline picture \cite{Levi:2020kvb}: 
(a) $0$-loop; 
(b) One-loop; 
(c), (e) Two-loop; 
(d), (f), (g) Three-loop.
Topology (e8) is a rank-two topology, whereas topologies (f5), (g4), and (g5), are rank-three 
topologies \cite{Levi:2020kvb}.}
\label{G4topo}
\end{figure}

In this paper we evaluate the contribution to the N$^3$LO sector that is quadratic in the 
spins, and originates from Feynman graphs at order $G^4$ in the diagrammatic expansion of the 
two-particle effective action in eq.~\eqref{2ptact}. A comprehensive analysis of the generic 
topologies at order $G^4$ in the EFT framework was presented in the recent work 
\cite{Levi:2020kvb}. Using the terminology defined in that work, these topologies are shown in 
figure \ref{G4topo}, classified according to their self-interaction vertices and the 
corresponding loop order in the worldline picture \cite{Levi:2020kvb}. As in 
\cite{Levi:2020kvb}, it is only the graphs that are at three-loop order in the worldline 
picture that genuinely represent a higher level of complexity in the N$^3$LO spinning sectors. 
These graphs have topologies (d), (f), or (g), as depicted in figure \ref{G4topo} (see also 
figure 12 in \cite{Levi:2018nxp}).

As of the two-loop order we also have higher-rank topologies, which means that more than a 
single basic integral of $n$-loop form, as specified in \cite{Levi:2020kvb}, is required to 
express integrals of that topology. In general, a rank-$r$ topology requires 
a combination of $r$ of the basic $n$-loop integrals (at order $G^{n+1}$). In this 
sector at order $G^4$ we have four higher-rank topologies: Topology (e8) is a rank-two 
topology, whereas topologies (f5), (g4), and (g5), are rank-three topologies 
\cite{Levi:2020kvb}. In contrast to the rank-one topologies, which easily boil down to 
one-loop computations, the higher-rank topologies require considerable processing mainly by 
means of integration-by-parts (IBP) \cite{Smirnov:2006ry}.

We proceed to the construction of the Feynman graphs that contribute to this sector. We recall 
that the spinning sector is more complicated than the non-spinning sector, because all 
possible topologies are realized at each order in $G$, even when the KK field decomposition is 
used \cite{Kol:2007bc,Levi:2008nh,Kol:2010ze}, in contrast to the non-spinning sector, where 
in particular at N$^n$LO for odd $n$, $n$-loop graphs are absent, as can be seen in table 
\ref{stateoftheart}. As we noted in section \ref{theoryquadspin} this sector contains three 
types of interactions of two distinct origins: There are the spin$_1$-spin$_2$ and 
spin$_1$-spin$_1$ interactions, which originate from the minimal coupling part of the action 
of a spinning particle in eq.~\eqref{spinptact}, and the interaction that involves the 
spin-induced quadrupole, which originates from the non-minimal coupling part of the action. 

Graphs with the first two types of interaction are readily constructed as subsets of the 
corresponding graphs in the N$^3$LO spin-orbit sector \cite{Levi:2020kvb} by replacing in the 
graphs of the latter some of the worldline mass couplings to the gravito-magnetic vector with 
its leading linear spin couplings. The presence of two spin couplings among the worldline 
insertions leads to fewer unique graphs than in the N$^3$LO spin-orbit sector, but still more 
than were present in the non-spinning sector. Graphs with the third type of interaction can 
similarly be obtained as a subset of the non-spinning N$^3$LO sector 
\cite{Levi:2011up,Foffa:2011ub} by replacing some of the worldline mass couplings to the 
scalar graviton with its leading spin-quadrupole coupling, or by adding a scalar graviton 
exchange to the few graphs at order $G^3$ that can be made static as a result (we remind the 
reader that this even-parity sector is static at order $G^4$).

\begin{figure}[t]
\centering
\includegraphics[angle=90,height=0.9\textheight,width=\textwidth]{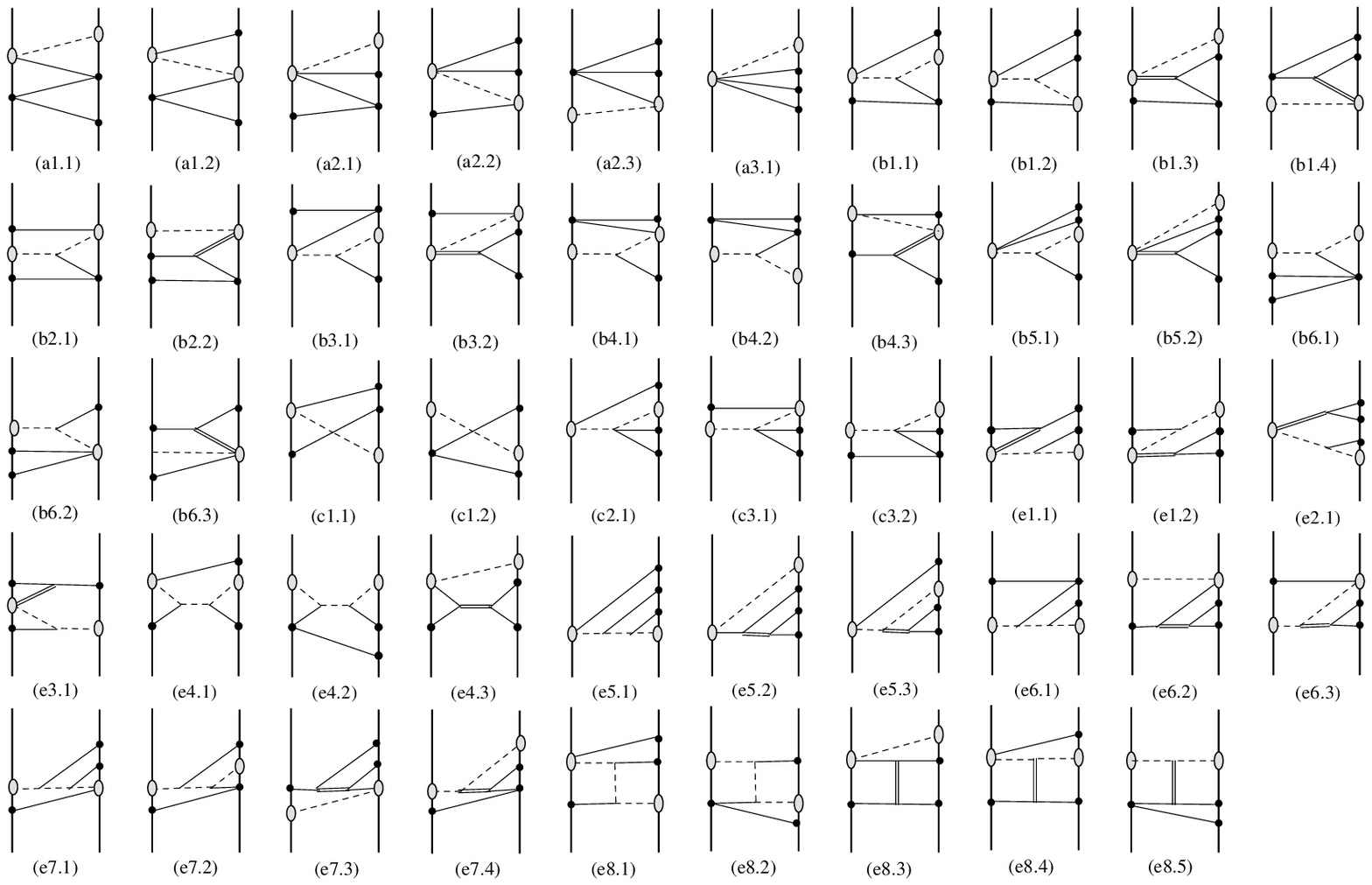}
\caption{Feynman graphs below three-loop order in the worldline picture, which contribute to 
the N$^3$LO spin$_1$-spin$_2$ static interaction at order $G^4$.
All of the graphs in this figure and the following ones should be accompanied by their 
`mirror' graphs, in which worldline labels are exchanged, namely $1\leftrightarrow2$.}
\label{below3loops1s2} 
\end{figure}

\begin{figure}[t]
\centering
\includegraphics[width=\textwidth]{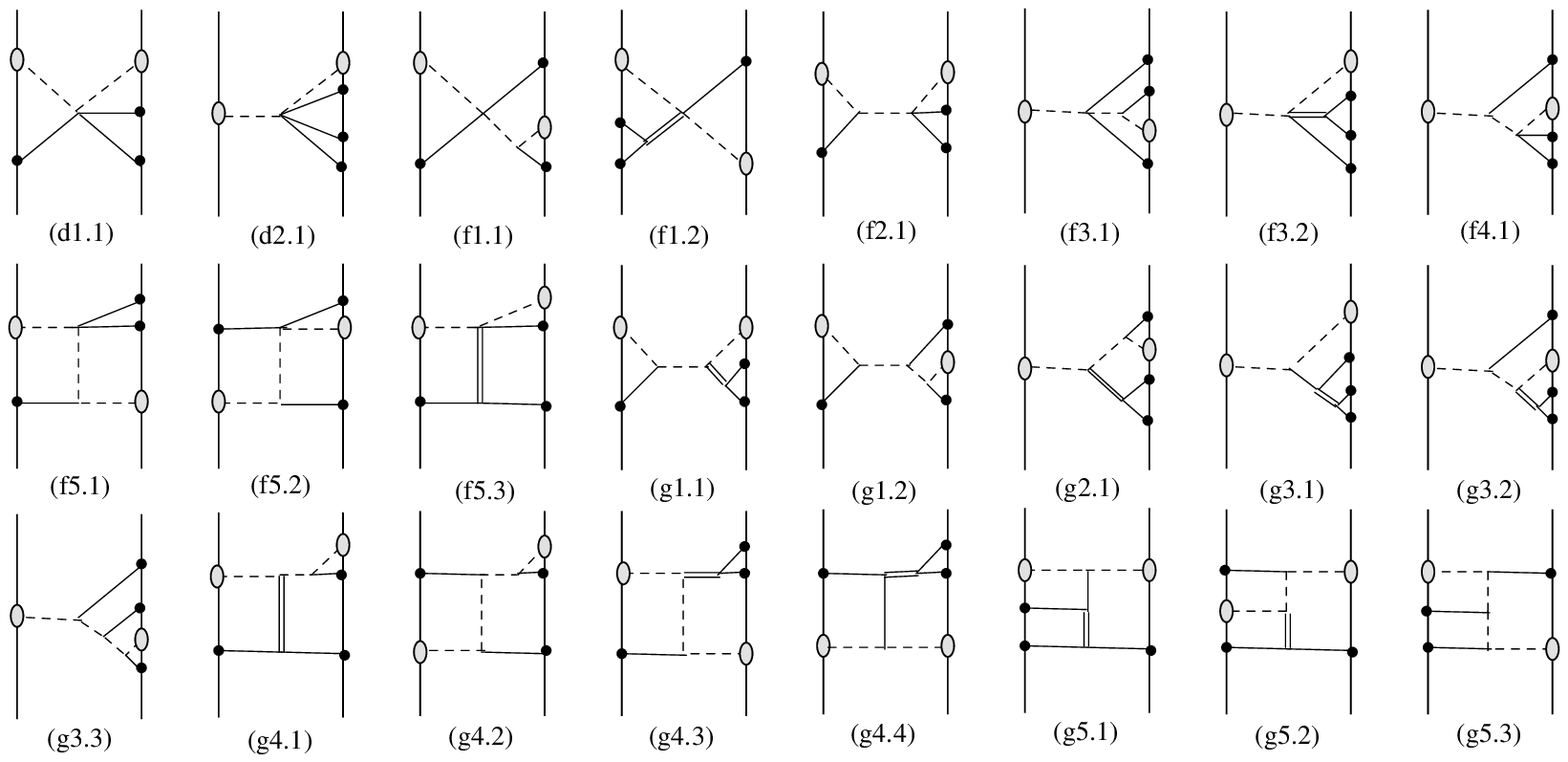}
\caption{Feynman graphs at three-loop order in the worldline picture, which contribute to 
the N$^3$LO spin$_1$-spin$_2$ static interaction at order $G^4$.}
\label{3loops1s2} 
\end{figure}

\begin{figure}[t]
\centering
\includegraphics[width=\textwidth]{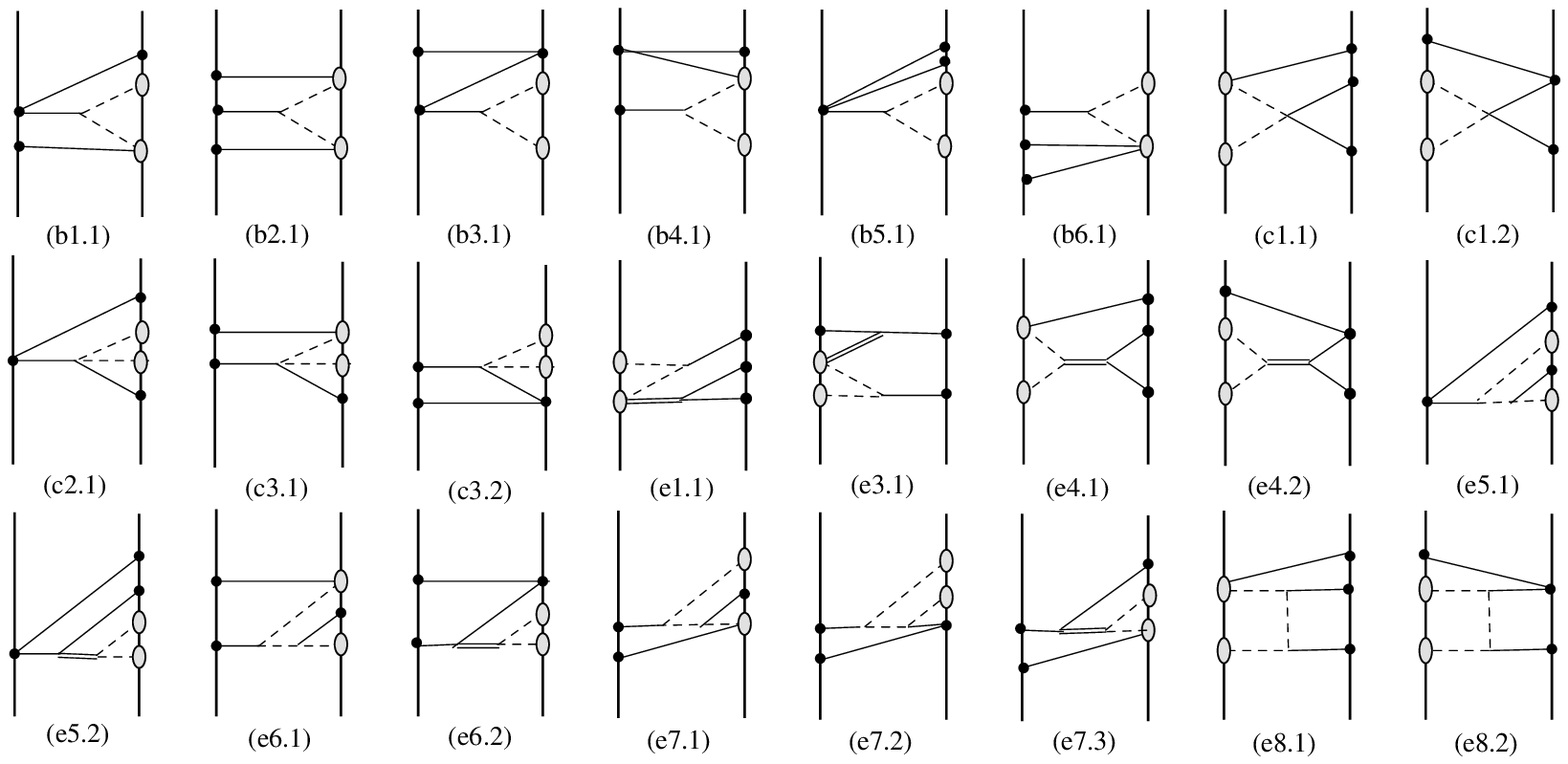}
\caption{Feynman graphs below three-loop order in the worldline picture, which contribute to 
the N$^3$LO spin$_1$-spin$_1$ static interaction at order $G^4$.}
\label{below3loops1s1} 
\end{figure}

\begin{figure}[t]
\centering
\includegraphics[width=\textwidth]{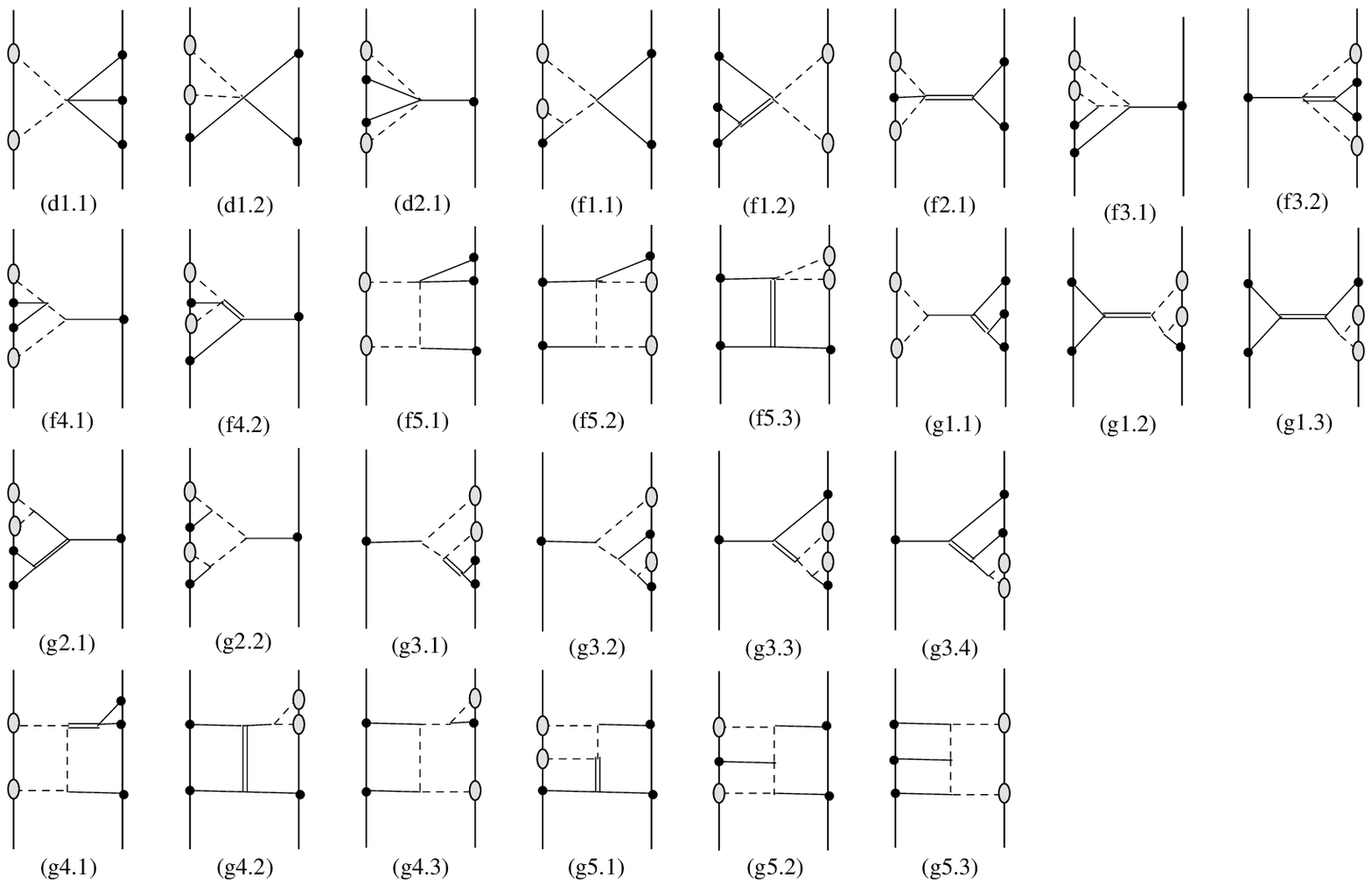}
\caption{Feynman graphs at three-loop order in the worldline picture, which contribute to 
the N$^3$LO spin$_1$-spin$_1$ static interaction at order $G^4$.}
\label{3loops1s1} 
\end{figure}

\begin{figure}[t]
\centering
\includegraphics[width=\textwidth]{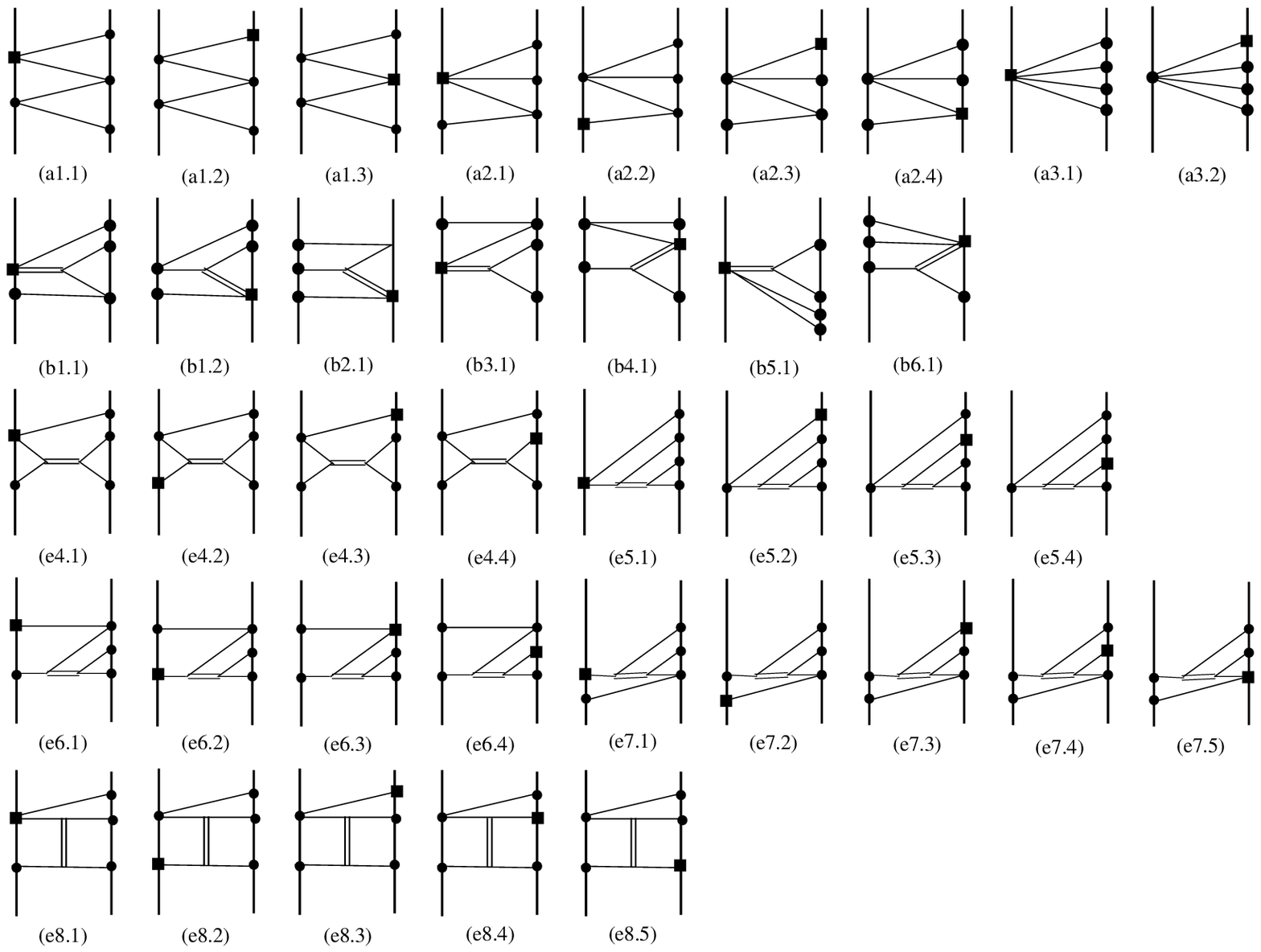}
\caption{Feynman graphs below three-loop order dependent on the spin-induced quadrupole, 
which contribute to the N$^3$LO quadratic-in-spin static interaction at order $G^4$. This is 
the only contribution to the sector with the spin-induced quadrupole.}
\label{below3loopssqquad} 
\end{figure}

Therefore, as no three-loop graphs enter at this order in the non-spinning case when the KK 
field decomposition is used, there are in fact no three-loop graphs with the spin-induced 
quadrupole which contribute to the N$^3$LO quadratic-in-spin sector; The only contribution to 
this sector dependent on the spin-induced quadrupole comes from graphs below three-loop order
in the worldline picture, as can be seen in figure \ref{below3loopssqquad}. Note that this is 
exactly what happens in the NLO spin-squared sector \cite{Levi:2015msa} due to the absence of 
one-loop graphs at the NLO non-spinning sector with the KK fields \cite{Kol:2007bc}. 
Altogether, this reasoning provided a complete crosscheck for the automated generation of the 
Feynman graphs using the \texttt{FeynGen} module of the \texttt{EFTofPNG} public code 
\cite{Levi:2017kzq}, which was extended to handle this sector. The resulting graphs are drawn 
in figures \ref{below3loops1s2}--\ref{below3loopssqquad} below (using JaxoDraw 
\cite{Binosi:2003yf,Binosi:2008ig} based on \cite{Vermaseren:1994je}). 

There are in total $163$ distinct graphs in this sector. Of these $52$ are three-loop graphs 
with spin$_1$-spin$_2$ and spin$_1$-spin$_1$ interactions, as shown in figures \ref{3loops1s2} 
and \ref{3loops1s1}, respectively. There are $31$ higher-rank graphs to evaluate. Of these 
$12$ are rank-two and $19$ are rank-three. The evaluation of the graphs was carried out using 
the upgrade to the publicly available \texttt{EFTofPNG} code which was described in 
\cite{Levi:2020kvb}, where the higher-rank graphs are reduced using projection 
\cite{Karplus:1950zza,Kniehl:1990iva,Binoth:2002xg} and IBP \cite{Laporta:2001dd} methods. Due 
to the additional spin with respect to the spin-orbit sector further projection of the 
numerators of the integrands was required. Further, more integrals show up in this sector 
compared to the corresponding spin-orbit one.


In the ancillary files to this publication we list the values of each of the 
Feynman graphs, both in PDF form and in machine-readable form.
We recall that all of the graphs in figures \ref{below3loops1s2}-\ref{below3loopssqquad} 
should be accompanied by their `mirror' graphs, in which worldline labels are exchanged, 
namely $1\leftrightarrow2$. We note that for the spin-squared interactions in figures 
\ref{below3loops1s1}-\ref{below3loopssqquad} we present the value for the graph with the 
spins on worldline ``1'', while for the spin$_1$-spin$_2$ interaction in figures 
\ref{below3loops1s2}-\ref{3loops1s2} we present the value for the graph with higher power of 
$m_1$. Two independent implementations of the framework outlined thus far were carried out in the 
present work in order to crosscheck and verify all the new results we present.

Similar features to those which were noted in \cite{Levi:2020kvb} are observed in this sector, 
with the new ones (which did not appear below N$^3$LO) similarly originating uniquely from 
three-loop topologies in the worldline picture. 
Let us enumerate the notable features encountered:
\begin{description}
\item [1.~Zeros.] There are $15$ graphs that vanish, 
$13$ of which have topologies (c1) and (e4): 
these vanish due to contact interaction terms, 
as discussed in \cite{Levi:2020kvb}. In addition, the 
graphs 4(e8.1), 4(e8.2) vanish, which was expected since these factorizable graphs contain 
the vanishing graph in figure 6(c1) in the N$^2$LO spin-squared sector \cite{Levi:2015ixa} as 
a subgraph.
\item [2.~Riemann zeta values.] There are $7$ graphs of the rank-three topologies 
(f5), (g4), (g5), that give rise to 
terms with $\zeta(2)\equiv \pi^2/6$, which occur as explained in \cite{Levi:2020kvb}.
\item [3.~Divergences and logarithms.] There are $36$ graphs, which make up the majority 
of the three-loop graphs, that give rise to simple poles in the dimensional parameter 
$\epsilon\equiv d-3$ in conjunction with logarithms in $r/R_0$, where $R_0$ is some 
renormalization scale. They appear in virtually all three-loop topologies (except (d2), (f3), 
(g2)) and are discussed in \cite{Levi:2020kvb}. 
\end{description}
We recall that the $\zeta(2)$ terms or the poles and accompanying logarithms originate 
independently from different basic integrals as explained in \cite{Levi:2020kvb}.

\section{\texorpdfstring{N$^3$LO}{N3LO} gravitational quadratic-in-spin action at 
\texorpdfstring{$G^4$}{G4}}
\label{result}

Summing all graphs at order $G^4$, we obtain the following contribution to the N$^3$LO 
gravitational quadratic-in-spin action :
\begin{align}
L_{\text{S$^2$}}^{\text{N$^3$LO}} & =
L_{\text{S$_1$S$_2$}}^{\text{N$^3$LO}} + 
L_{\text{S$_1^2$}}^{\text{N$^3$LO}} + 
L_{\text{C$_{1{ES^2}}$S$_1^2$}}^{\text{N$^3$LO}} 
+ \left( 1\leftrightarrow2 \right),
\end{align}
with
\begin{align}\label{n3los1s2}
L_{\text{S$_1$S$_2$}}^{\text{N$^3$LO}}
& =
-\frac{G^4}{r^6} \frac{1}{m_1m_2}\Bigg[\vec{S}_1\cdot\vec{S}_2
\Big(5 \, m_1^4 m_2 + 63\, m_1^3 m_2^2 \Big)
 -\vec{S}_1\cdot\vec{n}\vec{S}_2\cdot\vec{n}
\Big(25 \, m_1^4 m_2 + 300\, m_1^3 m_2^2 \Big)\Bigg],
\end{align}
\begin{align}\label{n3los1s1}
L_{\text{S$_1^2$}}^{\text{N$^3$LO}}
& =
\frac{G^4}{r^6} \frac{1}{m_1^2}\Bigg[S_1^2
\bigg(\frac{1}{14} \, m_1^4 m_2 -\frac{73}{70}\, m_1^3 m_2^2 - \, m_1^2 m_2^3\bigg)\nn\\
&
\qquad +(\vec{S}_1\cdot\vec{n})^2
\bigg(\frac{23}{7} \, m_1^4 m_2 + \frac{2851}{70}\, m_1^3 m_2^2 + 31\, m_1^2 m_2^3\bigg)\Bigg],
\end{align}
and
\begin{align}\label{n3loCES2}
L_{\text{C$_{1{ES^2}}$S$_1^2$}}^{\text{N$^3$LO}} 
& =
-\frac{G^4}{r^6} \frac{C_{1{ES^2}}}{m_1^2}
\bigg(S_1^2  - 3(\vec{S}_1\cdot\vec{n})^2\bigg)
\bigg(\frac{23}{28} \, m_1^4 m_2 + \frac{341}{14}\, m_1^3 m_2^2 
+ 57\, m_1^2 m_2^3 + 9\, m_1 m_2^4\bigg), 
\end{align}
where $\vec{r}\equiv\vec{r}_1-\vec{r}_2$ and $\vec{n}\equiv\vec{r}/r$.


Interestingly, in contrast to the situation in the N$^3$LO spin-orbit sector 
\cite{Levi:2020kvb}, all divergent terms, logarithms, and factors of $\zeta(2)$ that appear 
in the individual graphs leading to eqs.~\eqref{n3los1s2} and \eqref{n3los1s1} conspire to 
cancel out from the final result, which contains only finite rational terms. 
This is similar to the situation within analogous EFT derivations in the non-spinning sector, 
where all the poles in $\epsilon$, logarithms, and Riemann zeta values conspire to 
cancel out in each of the N$^n$LO sectors at $G^{n+1}$ as known for $n\leq5$ 
\cite{Foffa:2019hrb,Blumlein:2019zku}. 
We recall that both the non-spinning sector and the present one are static at the highest 
order in $G$, and have even parity with respect to the order of spin (see table 
\ref{stateoftheart}). This contrasts with the odd-parity spin-orbit sector, where the 
analogous piece is non-static. It is curious then whether this occurrence carries over to all 
of the even-parity sectors in spin. 

In any case, we know that the appearance and treatment of divergent terms and logarithms is 
related to the treatment of terms with higher-order time derivatives \cite{Blanchet:2013haa}. 
Since the appearance of higher-order time derivative terms is delayed by one order in the 
quadratic-in-spin sectors with respect to the spin-orbit sector, where it occurs already at 
the LO \cite{Levi:2010zu}, it may have been expected that the related novel features that show 
up in the final N$^3$LO spin-orbit result are absent from the final N$^3$LO quadratic-in-spin 
result here.

\section{Conclusions}
\label{theendmyfriend}

In this paper we derived for the first time the complete static N$^3$LO gravitational 
interactions that are quadratic in the spins in inspiralling compact binaries from Feynman 
diagrams with topologies at order $G^4$ within the framework of the EFT of spinning
gravitating objects \cite{Levi:2015msa}. The derivation builds on the recent work in 
\cite{Levi:2020kvb}, in which an upgrade of the \texttt{EFTofPNG} public code 
\cite{Levi:2017kzq} was carried out, with further extensions required for the present sector. 
The contribution we have calculated in this paper constitutes the most computationally 
challenging piece of the N$^3$LO quadratic-in-spin sector in terms of integration, due to the 
three-loop level that is the highest loop level in this sector. 
This sector enters at the $5$PN order for maximally-spinning compact objects, and complements 
the non-spinning sector at the same order, with relevant pieces in 
e.g.~\cite{Foffa:2019hrb,Blumlein:2019zku} and \cite{Foffa:2019eeb}.

This sector contains three types of interactions that originate from the two distinct parts of 
the effective action of a spinning particle, where in particular one interaction involves 
finite-size effects, which arise from the non-minimal coupling part of the point-particle 
action. Such effects distinguish between black holes and neutron stars, in contrast to the
spin-orbit coupling or point-mass interactions. Further, at this PN order higher non-minimal 
coupling operators beyond linear in the curvature with spin need to be included for the first 
time. We have correspondingly extended the effective action formulated in \cite{Levi:2015msa} 
to include operators that are quadratic in the curvature, similar to \cite{Levi:2020lfn} at 
the same PN order. While we find such operators that are quadratic in spin, and that can enter 
at this PN order, they do not enter at this order in $G^4$, and are thus left to a forthcoming 
publication. 

The N$^3$LO quadratic-in-spin interactions consist of a total of $163$ unique graphs at order 
$G^4$. Of these $52$ are three-loop graphs, which arise uniquely from the interactions that 
originate from minimal spin coupling, and in which special features are observed, similar to 
the analogous sector in the spin-orbit interaction \cite{Levi:2020kvb}. However, these
features (including divergences, logarithms, and transcendental factors) conspire to cancel 
out in the final result, similar to what happens in the non-spinning sector within EFT 
derivations for $n\leq5$ in the N$^n$LO sectors at order $G^{n+1}$ \cite{Blumlein:2019zku}. 
Nevertheless, the present sector is still considerably more challenging than the corresponding 
non-spinning sector. The evaluation of the remaining contributions up to order $G^3$ that are 
required to complete the N$^3$LO quadratic-in-spin sector should be facilitated with our 
extended upgrade of the \texttt{EFTofPNG} code \cite{Levi:2017kzq,Levi:2020kvb}, and this 
completion of the sector will be reported in forthcoming publications.

As previously noted, all of the sectors denoted with boldface entries in table 
\ref{stateoftheart}, including the current sector, have been derived to date for all generic 
only within the framework of the EFT of gravitating spinning objects 
\cite{Levi:2015msa,Levi:2017kzq,Levi:2019kgk,Levi:2020kvb}. 
Clearly, independent crosschecks of these precision results are essential, and thus 
overlapping studies, possibly with further modern amplitudes methods as in 
\cite{Cachazo:2017jef,Guevara:2017csg,Cheung:2018wkq,Bern:2019nnu,Bern:2019crd}, 
are very welcome.

\acknowledgments

ML receives funding from the European Union's Horizon 2020 research and 
innovation programme under the Marie Sk{\l}odowska Curie grant agreements 
No.~847523 and No.~764850, and from the Carlsberg Foundation. 
ML has also been supported by the European Union's Horizon 2020 Framework 
Programme FP8/2014-2020 ``preQFT'' starting grant No.~639729. 
ML is grateful to Freddy Cachazo for the warm hospitality at Perimeter 
Institute where the final stages of this work were carried out.
AJM and MvH are both supported by an ERC starting grant No.~757978 and 
a grant from the Villum Fonden No.~15369. 
AJM is also supported by a Carlsberg Postdoctoral Fellowship (CF18-0641).
MvH is also supported by the European Union's Horizon 2020 research and 
innovation programme under grant agreement No.~793151.



\bibliographystyle{jhep}
\bibliography{gwbibtex}

\end{document}